\documentclass[letterpaper, 10 pt, conference]{ieeeconf}  

\IEEEoverridecommandlockouts                              

\usepackage{graphics} 
\usepackage{epsfig} 
\usepackage{mathptmx} 
\usepackage{times} 
\usepackage{amsmath} 
\usepackage{amssymb}  
\usepackage{eucal}
\usepackage{algorithm2e}
\usepackage{algorithmic}
\usepackage{multirow}
\usepackage{yhmath}
\usepackage{color}
\usepackage{graphicx}
\usepackage{epstopdf}
\usepackage{indentfirst}

\title{\LARGE \bf
Towards Provably Safe Mixed Transportation Systems with Human-driven and Automated Vehicles
}

\author{Xi Liu$^{1}$, Ke Ma$^{1}$, and P. R. Kumar$^{1}$
\thanks{*This material is based upon work partially supported by NSF under Contract Nos. CPS-1239116 and Science $\&$ Technology Center Grant CCF-0939370, and NPRP grant 6-784-2-329 from the Qatar National Research Fund, a member of Qatar Foundation.}
\thanks{$^{1}$Xi Liu, Ke Ma and P. R. Kumar are with the Department of Electrical and Computer Engineering,
        Texas A$\&$M University, 
        {\tt\small $\{$xiliu,prk,ke.ma$\}$@tamu.edu}}%
}

\begin{document}

\maketitle
\thispagestyle{empty}
\pagestyle{empty}

\begin{abstract}

Currently we are in an environment where the fraction of automated vehicles is negligibly small. We anticipate that this fraction will increase in coming decades before, if ever, we have a fully automated transportation system. Motivated by this we address the problem of provable safety of mixed traffic consisting of both intelligent vehicles (IVs) as well as human driven vehicles (HVs). An important issue that arises is that such mixed systems may well have lesser throughput than all human traffic systems, if the automated vehicles are expected to remain provably safe with respect to human traffic. This necessitates the consideration of strategies such as platooning of automated vehicles in order to increase the throughput. In this paper we address the design of provably safe systems consisting of a mix of automated and human-driven vehicles including the use of platooning by automated vehicles.

We design motion planing policies and coordination rules for participants in this novel mixed system. HVs are considered as nearsighted and modeled with relatively loose constraints, while IVs are considered as capable of following much tighter constraints. HVs are expected to follow reasonable and simple rules. IVs are designed to move under a model predictive control (MPC) based motion plans and coordination protocols. Our contribution of this paper is in showing how to integrate these two
types of models safely into a mixed system. System safety is proved in single lane scenarios, as well as in multi-lane situations allowing lane changes. 

\end{abstract}

\section{INTRODUCTION}

There has been much interest in intelligent vehicles (IV), evidenced by the ITS program in the US, the EUREKA Prometheus Project in the European Union, and the ITS initiative program in Japan. A Google self-driving car won the 2005 DARPA Grand Challenge, and a Toyota Prius modified with Google's self-driving technology was licensed in Nevada State. Motivated by this, there has been research on automated transportation aimed at developing control laws and protocols that result in safety and liveness of the traffic system [1-3]. 

In order to attain such a fully automated transportation system starting from where we are currently, viz., a human-driven transportation system, we will need to address the problem of \emph{transition} where there is traffic consisting of a mix of human-driven and automated vehicles. At the present moment, the fraction of IVs among all vehicles is negligibly small. As time progresses, if the proportion of automated vehicles in the mix is to become larger, then we will necessarily need solutions for automated vehicles that are provably safe in the presence of varying percentages of human-driven traffic. 

When considering automated transportation in a mixed environment, another important issue arises. If one wants a provably safe system, then the throughput of automated vehicles may in fact turn out to be less than the throughput of a purely human driven system. The reason is that humans do not drive in a provably safe manner. They dangerously trade off safety for throughput, such as when following cars too closely. Hence, in order to improve throughput, we will need to specifically employ strategies that can improve the throughput beyond that of human driven systems. One such strategy is ``platooning'' which was proposed in [9-10] and tested in an implementation in [11-12].

In this paper we address the issue of increased throughput as well as safety in mixed systems consisting of both human and automated vehicles. While the automated vehicles resort to advanced strategies they need to however respect the margins needed by human drivers, such as allowing for the slow reaction to braking of vehicles ahead of them, and other such considerations. That is the goal of this paper. We aim to establish system wide safety, i.e., safety of the entire system, comprised of arbitrary proportions of automated and human-driven vehicles (HVs), with the automated vehicles pursuing strategies such as platooning in order to increase throughput. 

Generation of collision-free trajectories for IV motion planning has been widely researched, e.g., the Monte Carlo approach [4], reachable set analysis [5] and prediction of potential crash behaviors [6]. Typically, these results, as in [4,6], do not address provable safety. There is limited literature incorporation of human-driven vehicles from a system design view. Rajeev et al. [7] consider the presence of HVs in such a system by modeling them as game participants with imperfect environment state information compared to IVs with perfect information. Au et al. [8] design a mixed traffic intersection compatible with IVs and HVs with driver assistant systems. Our design follows the approach of the work in [3], and is apparently the first to address provable safety of automated traffic in a mixed environment.

In Section II, we establish safety for single lane traffic. We consider a model predictive control (MPC) motion planner for IVs. Even though such MPCs consider only finite time behavior, we establish safety which is an infinite time property. We next analyze vehicles in multi-lane traffic, design easy-to-follow rules for HVs, and develop new MPC and coordination protocols for IVs to follow other vehicles in Section III. We also prove safety of the automated vehicles with respect to the other vehicles based on the proposed protocols and rules for multi-lane traffic in Section III. A simulation study of proposed system is provided in Section IV. Concluding remarks follow in Section V.
\section{SINGLE LANE SAFETY}
We consider three elements in the mixed system: manually-driven vehicles (called HVs), automated vehicles (called IVs) and platooning by a group of electronically connected IVs with small inter-vehicle separations. We assume IVs communicate with other IVs via V2V communication (DSRC, IEEE 802.11p) [11] and exchange intentions with HVs by reading each other's onboard signal lights.

  \begin{figure}[thpb]
      \centering
      \includegraphics[scale=0.3]{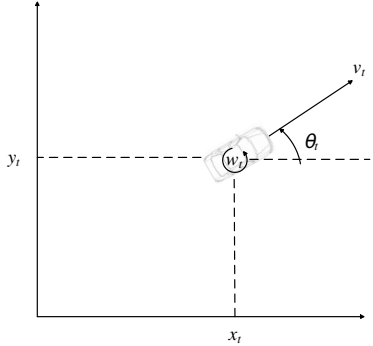}
      \caption{The Unicycle Kinematic Vehicle Model}
   \end{figure}

We model vehicles by unicycle kinematic models, as shown in Fig. 1. At time $t$, the state information of vehicle $c$ is $ \mathbf{x}_{t}(c):=(x_{t}(c), y_{t}(c), \theta_{t}(c))^{T}$ capturing its position and orientation. Where there is no scope for confusion we omit the vehicle identifier $c$. The input vector is $\mathbf{u}_{t}:=(v_{t},\omega_{t})$ denoting velocity control and steering control. We assume sampled data and control signals are maintained constant during a time slot $[t,t+h)$. There are physical constraints on the input signals: (1) $v_{t}\in[0,v_{max}]$; (2) $\theta_{t}\in[\theta_{min},\theta_{max}]$; (3) $\varDelta{v_{t}}\in[a_{min}h,a_{max}h]$. The quantity $a_{min}<0$ is the maximum achievable (i.e., most rapid) deceleration. We allow for different capabilities for human and intelligent vehicles; we use a superscript ``h'' and ``i'' to differentiate them. Specifically, we allow $a^h_{min} > a^i_{min}$, i.e., human driven vehicles  can brake less strongly than intelligent vehicles. The kinematic equation of a vehicle is: 
\begin{align}
\mathbf{x}_{t+h}:=f(\mathbf{x}_{t}, \mathbf{u}_{t}),
\end{align}
where, if $\omega_{t}\neq0$:
\begin{align*}
x_{t+h}:&=2\dfrac{v_{t}}{\omega_{t}}\sin(\dfrac{1}{2}\omega_{t}h)\cos(\theta_{t}+\dfrac{1}{2}\omega_{t}h)+x_{t}
\\
y_{t+h}:&=2\dfrac{v_{t}}{\omega_{t}}\sin(\dfrac{1}{2}\omega_{t}h)\sin(\theta_{t}+\dfrac{1}{2}\omega_{t}h)+y_{t}
\\
\theta_{t+h}:&=\omega_{t}h+\theta_{t},
\end{align*}
while if $\omega_{t}=0$:
\begin{align*}
x_{t+h}:&=v_{t}h\cos(\theta_{t})+x_{t}
\\
y_{t+h}:&=v_{t}h\sin(\theta_{t})+y_{t}
\\
\theta_{t+h}:&=\theta_{t}.
\end{align*}

We begin by consider mixed traffic, i.e., consisting of both HVs and IVs, driving along a single lane as shown in Fig. 2. Our design approach in this simple scenario is to ensure that each IV is responsible for not colliding with the vehicle in front of it, and, at the same time, behaving in a manner similar to an HV when followed by HVs. Thereby, we ensure that HVs do not need to differentiate whether a lead vehicle is an IV or HV. However, an IV must ensure that if a following vehicle is an HV, then it can only brake with $a^{h}_{min}$ to avoid collision. The main contribution of this section is to show how to safely integrate such a loose model of an HV and a tight model of an IV in a mixed system.
\begin{figure}[thpb]
      \centering
      \includegraphics[scale=0.38]{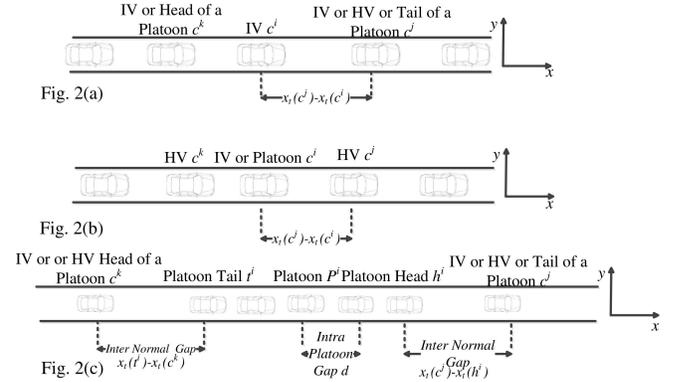}
      \caption{Single Lane Traffic in Mixed System}
\end{figure}

We will allow IVs to form ``platoons'' [12-13]. A platoon is a set of vehicles separated by small distances, with all moving together as a formation; see Fig. 2(b)-(c). Such a platoon can be treated as a single ``long'' IV, with the head of a platoon responsible for not colliding with the vehicle in front of the platoon. Within each platoon, each IV is responsible for not colliding with its lead vehicle. Feedback control with state information of head vehicle can be applied to maintain string stability [14]. We note that if the platoon is followed by an HV, then all agents within the platoon can only decelerate with $a^{h}_{min}$.

\subsection{Safety with only IVs} 
To build up to mixed traffic, we first consider the case where there are only IVs on the single lane. For single lane traffic, since there is no need for steering to change lane, we ignore angle and orientation issues, and simply suppose that the state is one-dimensional and indicates the distance along the lane, $x_{t+h}(c)=f(x_{t}(c), v_{t}(c))=x_{t}(c)+v_{t}(c)h$. As shown in Fig. 2(a)-(b), let $c^{i}$ be the IV we are interested in, $c^{j}$ the vehicle $c^{i}$ is following, and $c^{k}$ the vehicle that $c^{i}$ is followed by. Denote by $C_{I}$ the set of all IVs. The position and velocity of the nearest lead vehicle $c^{j}$ is
\begin{align}x_{t}^{Lead}:=x_{t}(c^{j})\\
v_{t}^{Lead}:=v_{t}(c^{j}),\nonumber
\end{align}
where $c^{j}:=\arg\underset{c\in C}{\min}\,\{x_{t}(c):x_{t}(c^{i})<x_{t}(c)\}$. Given $v_{t-h}$ at time $t$, we can estimate the range of $v_{t}$ based on physical constraints as $v_{t}\in[\underline{v_{t}}, \overline{v}_{t}]$, where:
\begin{align}
\underline{v}_{t}:=\max\{0, v_{t-h}+a_{\min}h\}
\end{align}
\begin{align}
\overline{v}_{t}:=\min\{v_{max},v_{t-h}+a_{max}h\}.
\end{align}

We consider the following MPC to govern the movement of an IV in single lane, which generalizes a result of [3] by taking into account the lead vehicle's velocity:

\leftline{\textbf{MPC for IVs in Single Lane}}
\begin{align}
\underset{\mathbf{u}(0:N-1)}{\min}\,\hspace{8mm} &J(x_{t}, x_{t}^{f}, {\mathbf{u}(0:N-1)})\\
s.t.\hspace{8mm} &x_{t+(k+1)h}^{Lead}-x_{t+(k+1)h}\geqslant{D^{0}_{s}(v_{t+kh},v_{t+kh}^{Lead},a_{min})}\nonumber\\
&x_{t+(k+1)h}^{Lead}=x_{t+kh}^{Lead}+(v_{t+(k-1)h}^{Lead}+a_{min}h)h\nonumber\\
&x_{t+(k+1)h}=x_{t+kh}+v_{t+kh}h\nonumber\\
&a_{min}h\leqslant\Delta{v_{t+kh}}=v_{t+kh}-v_{t+(k-1)h}\leqslant{a_{max}h}\nonumber\\
&v_{t+kh}\in[{\underline{v}_{t+kh}}, \overline{v}_{t+kh}]\nonumber
\end{align}
for all $k\in\{0,...,N-1\}$, where, $\mathbf{u}(0:N-1)=\{v_{t},...,v_{t+(N-1)h}\}$, and $N$ is the length of the time horizon $J$ in the objective function for this MPC. The cost function $J$ is allowed to be arbitrary, since our guarantee of safety in the following Theorem depends only on the existence of a feasible solution to the MPC, and not on the objective function or its value. The distance between vehicles in a platoon is
\begin{align}
D^{0}_{s}(v_{t-h},v_{t-h}^{Lead},a_{min}):=
&\dfrac{v_{t-h}^2-v_{t-h}^{2Lead}}{-2a_{min}}
\\&+(v_{t-h}-v_{t-h}^{Lead})h-\dfrac{1}{2}a_{min}h^{2}+d_{min},\nonumber
\end{align}
where $d_{min}\geq0$ is the minimal gap set up beforehand. Different from [3], whose smallest following distance grows polynomially with velocity $d_{s}(v_{t-h},a_{min})=v_{t-h}^2/-2a_{min}+v_{t-h}h-a_{min}h^{2}/2>v_{t-h}^2/-2a_{min}$, our following distance $D^{0}_{s}(v_{t-h},v^{Lead}_{th},a_{min})$ can be as small as $d_{min}-a_{min}h^{2}/2$, independent of velocity. This property makes the MPC amenable to platooning. Below theorem can be proved by checking once initial condition is satisfied, there always exits deceleration with $a_{min}$ as a feasible solution for MPC for any following vehicle, which is omitted due to page limitation.

\noindent\textbf{Theorem 1.} Suppose at any initial time $t$, one has $x_{t}^{Lead}-x_{t}\geqslant{D^{0}_{s}(v_{t-h},v_{t-h}^{Lead},a_{min})}$ for every pair $c^{i}, c^{j}\in{C_{I}}$ with $x_{t}(c^{i})<x_{t}(c^{j})$ and $C_{H}=\emptyset$. Then the single lane system is safe if all IVs move under the above MPC motion planning control.
\subsection{Integration of HVs}
In this subsection, we consider mixed traffic consisting of HVs and IVs together in a single lane. Let $C_{I}$ denote the set of IVs, and by $C_{H}$ the set of HVs. We suppose that HVs follow a lead vehicle with separation distance larger than $D^{0}_{s}$, since they have $a^{h}_{min}>a^{i}_{min}$. We prove that system safety is guaranteed if the HVs conform to this rule when the IVs move under MPC (5) with different following distances. 

By Theorem 1, we have safety if HVs follow the designed rule when their lead vehicle is also an HV. It is also safe if an IV, followed by another IV, is following an HV, because, the last IV effectively sees lesser acceleration passed on to it through the interposed IV. A solution feasible for preparing the HV to decelerate with looser deceleration is always feasible for preparing the HV to decelerate with tighter deceleration. However, it is unsafe if an IV $c^{i}$ still follows (5) when following any vehicle $c^{j}$ but is followed by an HV $c^{k}$, because $c^{k}$ presumes $c^{i}$ will not decelerate with deceleration less than $a^{h}_{min}$. Consequently, a new following distance is desiasgned below for IVs in such cases, in which $c^{i}$ can brake with $a^{h}_{min}$ to avoid collision with $c^{j}$.

Consider, as shown in Fig. 2, $c^{i}\in{C_{I}}$ followed by $c^{k}\in{C_{H}}$, and following $c^{j}\in{C}$. If $c^{i}$ can only decelerate with $a^{h}_{min}>a^{i}_{min}$, then to avoid collision with $c^{j}$, we define a more stringent following distance for $c^{i}$ in the following theorem. The position and velocity of the nearest following vehicle $c^{k}$ is:
\begin{align} x_{t}^{Follow}:=x_{t}(c^{k})\\
v_{t}^{Follow}:=v_{t}(c^{k}),\nonumber
\end{align}
where $c^{k}:=\arg\underset{c\in C}{\max}\,\{x_{t}(c):x_{t}(c^{i})>x_{t}(c)\}$.

\noindent\textbf{Rule for HVs in Single Lane Traffic:}

\noindent We suppose that, HVs follow their lead vehicles with separation distance larger than $D^{0}_{s}(v_{t-h},v^{Lead}_{t-h},a^{h}_{min})$, but do not take into consideration whether there are any vehicles following them.

\noindent\textbf{Theorem 2.} Suppose the following conditions are satisfied at an initial time $t$, (i) $x_{t}^{Lead}-x_{t}\geqslant{D^{0}_{s}(v_{t-h},v_{t-h}^{Lead},a^{i}_{min})}$ and $x_{t}-x^{Follow}_{t}\geqslant{D^{0}_{s}(v^{Follow}_{t-h},v_{t-h},a^{h}_{min})}$ for every $c^{k}\in{C_{I}}, c^{i}\in{C_{I}}, c^{j}\in{C}$ (ii) $x_{t}^{Lead}-x_{t}\geqslant{D^{1}_{s}(v_{t-h},v_{t-h}^{Lead})}$ and $x_{t}-x^{Follow}_{t}\geqslant{D^{0}_{s}(v^{Follow}_{t-h},v_{t-h},a^{h}_{min})}$ for every $c^{k}\in{C_{H}}, c^{i}\in{C_{I}}, c^{j}\in{C}$.
Then the single lane system with IVs and HVs is safe if all HVs conform to the rule above, and IVs move under any feasible solutions of the MPC (5) motion planning control with right-hand side constraints replaced by $D^{1}_{s}(v_{t-h},v^{Lead}_{t-h})$ and $D^{0}_{s}(v_{t-h},v^{Lead}_{t-h},a^{i}_{min})$, where:
\begin{align}
D^{1}_{s}(v_{t-h},v^{Lead}_{t-h})&:=\dfrac{v_{t-h}^2}{-2a^{h}_{min}}-\dfrac{v_{t-h}^{2Lead}}{-2a^{i}_{min}}+(v_{t-h}-v_{t-h}^{Lead})h\\&-\dfrac{3}{2}(a^{h}_{min}-a^{i}_{min})h^{2}\dfrac{v^{Lead}_{t-h}}{-a^{i}_{min}h}-\dfrac{a^{h}_{min}h^{2}}{2}+d_{min}.\nonumber
\end{align}
\subsection{Integration of Platooning and Diverse IVs}

We now turn to the problem of automated platooning. A platoon is a contiguous group of IVs, with small inter-vehicular separation, with all vehicles moving at almost the same velocity and acceleration/deceleration. The motivation for using platooning is that a small gap increases traffic capacity [9-13]. There are three maneuvers needed for platooning in single lane traffic: platoon join, platoon split and platoon maintain. All maneuvers must be executed while maintaining inter-platoon and intra-platoon safety. Thus we retain all the earlier constraints in MPC motion planning for platoons. Different objective functions are designed to execute different maneuvers as listed in Table I, where $x_{t+kh}^{f}$ is the target position set for time $t+kh$, $d^{f}_{k}$ is the target spacing set for time $t+kh$, $d$ is the intra-platoon spacing, and $\alpha$ is discount factor. The reason for designing discount factor is that, although the result from $t$-round MPC (the MPC whose initial time is $t$) is a plan for a time horizon, we only implement it for the first time slot $[t,t+h)$. Hence, in one round MPC target spacing is expected to be achieved the earlier the better. We note from the constraints in the MPC that $d\geq{a_{min}h^{2}/2+d_{min}}$ implicitly has a lower bound, and from the nature of the split $d^{f}_{k}$ grows with $k$.

\begin{table}[h]
\caption{Objective Functions for Different Maneuvers}
\label{table_example}
\begin{center}
\begin{tabular}{|c||c|}
\hline
Maneuver & Objective Function\\
\hline
Follow& $J=\sum_{k=0}^{N-1}(x_{t+kh}-x_{t+kh}^{f})^{2}$\\
\hline
Join/Maintain& $J=\sum_{k=0}^{N-1}(x_{t+kh}-x_{t+kh}^{Lead}-d)^{2}e^{-\alpha{k}}$ \\
\hline
Split& $J=\sum_{k=0}^{N-1}(x_{t+kh}-x_{t+kh}^{Lead}-d^{f}_{k})^{2}e^{-\alpha{k}}$\\
\hline
\end{tabular}
\end{center}
\end{table}

In single lane traffic, a platoon is formed by consecutive IVs. A ``free agent'' IV can join a platoon at its tail or head. Also, a platoon may need to split to release one of its vehicles and allow it to become a free agent. When a free agent executes a join maneuver at the tail of a platoon, its objective function changes from ``following'' to ``join'' while the head of the platoon needs no adjustment. If the join happens at the head, then the head of the platoon changes its objective function. After a join, only the subsequent head generates a route for the whole platoon, while the others tightly follow their lead vehicle one by one. Similar considerations apply to a split.

\section{MULTI-LANE SAFETY}
We next consider mixed traffic driving along a multi-lane road. In this case the new complication that arises is lane changing by vehicles.
For a vehicle $c$ at time $t$, we denote by $\alpha_{t}(c)$ the vehicle $c$'s target lane, and  by $\beta_{t}(c)$ its current lane. Each vehicle has three states: $free$ (without a lane change request), $wait$ (with a lane change request but waiting for a safe time to initiate lane change) and $processing$ (changing lane). We denote by $s(t-h)$ the state during time slot $[t-h,t)$, with the state being constant during the entirety of a time slot. IVs know each other's states via periodic broadcasts, and read HVs' intentions from their turn lights. Fig. 3 illustrates the state transition diagram of IVs and HVs. Both IVs and HVs are allowed to generate a new $\alpha_{t}(c)$ at any time after finishing one instance of lane change; as shown in he state transition for both from the state $free$ to $wait$ in Fig. 3. Their differences, as will be shown, are (i) they pose different safety concerns to followers, (ii) more strict constraints should be satisfied for IVs to initiate lane change than for HVs. The difference (ii) is embodied in the following design of safety-related sets. A safety related set is a set of vehicles that the vehicle under consideration must pay attention to while planning its movement.

\begin{figure}[thpb]
      \centering
      \includegraphics[scale=0.26]{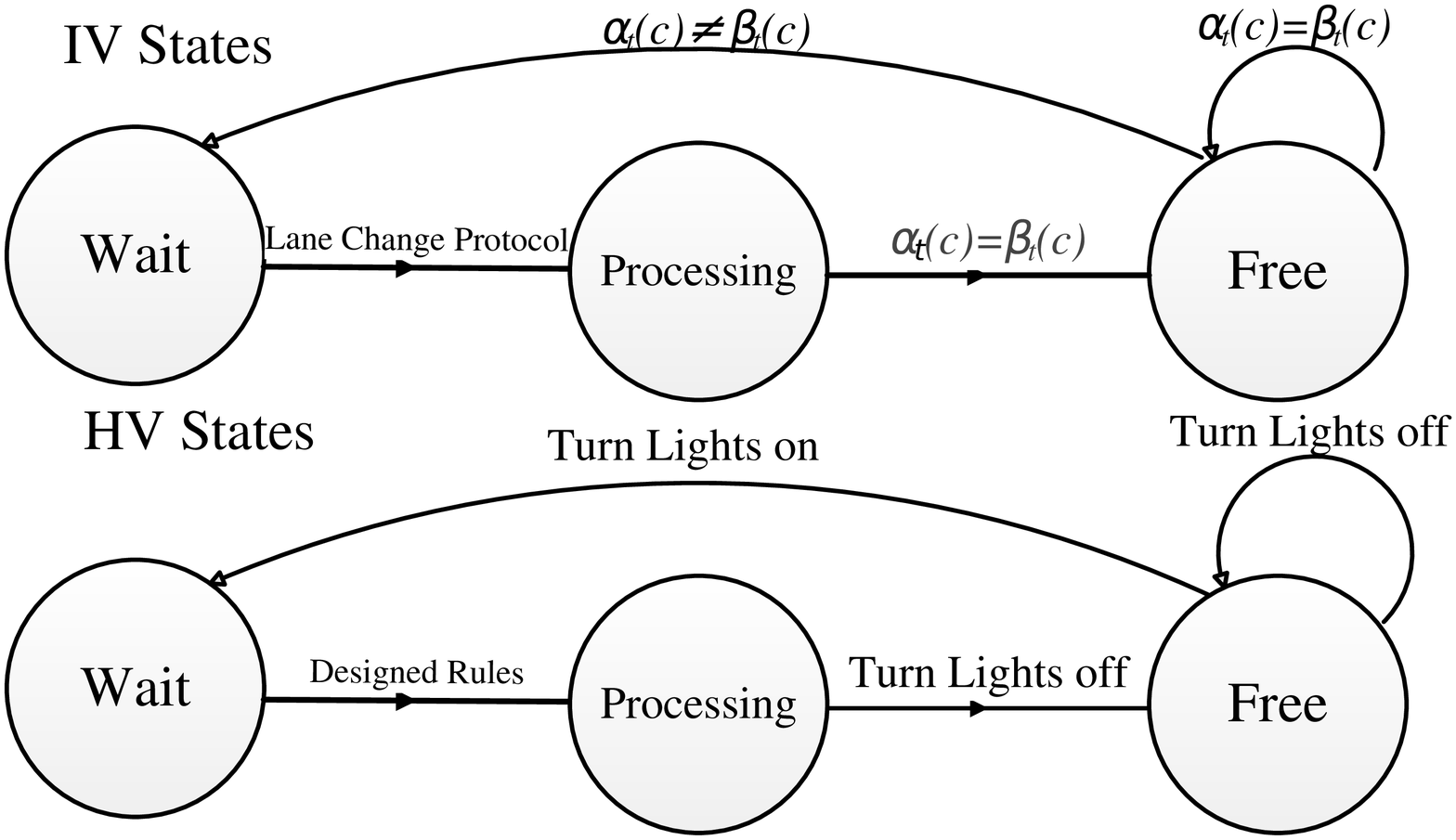}
      \caption{IV States and HV States}
\end{figure}

We first consider safety-related sets from the view point of an IV $c^{i}$ which is in state $wait$ at time $t$, and seeking a safe time to change lane. To define safety-related sets, we must extend the definition of ``follow'' in multi-lane traffic: a vehicle $c$ may need to follow another vehicle not in its lane by behaving as if there is an imaginary vehicle in $c$ 's current lane, which has the same velocity and $x$ coordinate as the vehicle in another lane. Also, we refer to vehicles whose $x$ coordinate is smaller or bigger than $c$ at time $t$ in different lanes by using the term vehicles ``ahead of'' or ``behind'' $c$. We denote by (i) $C^{+}_{I_{1}}(t)$ the set of all IVs in state $free$ ahead of $c^{i}$ which are in lane $\alpha_{t}(c^{i})$ or $\beta_{t}(c^{i})$, (ii) $C^{+}_{I_{2}}(t)$ the set of all IVs ahead of $c^{i}$ in state $wait$ or $processing$ which plan to change lane from or to $\beta_{t}(c^{i})$ or $\alpha_{t}(c^{i})$ at time $t$, (iii) $C^{+}_{H}(t)$ the set of all HVs ahead of $c^{i}$ at time $t$. As will be designed in the lane change protocol, $c^{i}$ ensures a following distance of at least $D^{0}_{s}(v_{t-h},v^{Lead}_{t-h},a_{min})$ for vehicles in $C^{+}_{I_{1}}(t)$, which is smaller than $d_{s}(v_{t-h},a_{min})$ for vehicles in $C^{+}_{I_{2}}(t)\cup C^{+}_{H}(t)$. The reason is that one can prove that $D^{0}_{s}(v_{t-h},v^{Lead}_{t-h},a_{min})$ is enough for $c^{i}$ to ensure safety of vehicles ahead of it if they do not change lane during $c^{i}$'s lane change:
\begin{align} 
C^{+}_{I_{1}}(t):=\{c^{j}\in{C_{I}\backslash{c^{i}}}:(x_{t}(c^{i})\leqslant{}x_{t}(c^{j}))\\
\wedge(\beta_{t}(c^{j})=\beta_{t}(c^{i})\lor{\alpha_{t}(c^{i})})\wedge(s(t-h)=free)\},\nonumber\\
C^{+}_{I_{2}}(t):=\{c^{j}\in{C_{I}\backslash{c^{i}}}:(x_{t}(c^{i})\leqslant{}x_{t}(c^{j}))\\
\wedge(\alpha_{t}(c^{j})\lor{\beta_{t}(c^{j})}=\beta_{t}(c^{i})\lor{\alpha_{t}(c^{i})})\nonumber\\
\wedge(s(t-h)=wait\lor{processing})\},\nonumber\\
C_{H}^{+}(t):=\{c^{j}\in{C_{H}}:x_{t}(c^{i})\leqslant{}x_{t}(c^{j})\}.
\end{align}
To ensure that during $c^{i}$'s lane change vehicles ahead of $c^{i}$ do not change lane from an unsafe distance, we require that any such vehicle $c^{j}$ must take into consideration the set $C^{-}(t)$ consisting of all IVs behind it which are currently in or changing from or to $\beta_{t}(c^{j})$ or $\alpha_{t}(c^{j})$, which is denoted by $C_{I}^{-}(t)$. To ensure that $c^{i}$'s lane change will not pose a safety threat to any HVs behind it, we require that before lane change $c^{i}$ must consider the set $C_{H}^{-}(t)$ of all HVs behind $c^{i}$:
\begin{align} 
C_{I}^{-}(t):=\{c^{k}\in{C_{I}}\backslash{c^{i}}:(x_{t}(c^{i})\geqslant{}x_{t}(c^{k}))\\
\wedge((\beta_{t}(c^{k})=\beta_{t}(c^{i})\lor{\alpha_{t}(c^{i})})\nonumber\\
\lor{((\alpha_{t}(c^{k})=\beta_{t}(c^{i})}\lor{\alpha_{t}(c^{i}))\wedge{(s(t-h)=processing)}))}\}\nonumber,\\
C_{H}^{-}(t):=\{c^{k}\in{C_{H}}:x_{t}(c^{i})\geqslant{}x_{t}(c^{k})\}.
\end{align}
The vehicle $c^{i}$ does not take into account IVs which are in state $wait$ and are seeking a safe time to change lane to $\beta_{t}(c^{i})$ or $\alpha_{t}(c^{i})$. The reason is that for any such vehicle $c^{k}$, since $c^{i}$ is in state $wait$, it is a lead vehicle in ${C^{+}_{I_{2}}(t)}$ with respect to $c^{k}$. Hence $c^{k}$ will not transit to $processing$ before it ensures that the lane change stays safe with $c^{i}$ irrespective of $c^{i}$'s behavior. 

The safety related sets discussed above for IVs in state $wait$ are used to decide when is a safe time to initiate lane change. For IVs in state $wait$ and $free$, since we set $\alpha_{t}(c^{j})\neq \beta_{t}(c^{i})$ until the lane change of any vehicle $c^{j}$ is completed, those IVs sometimes need to follow vehicles in state $processing$, which are not in their lane. For an IV $c^{i}$ which is in state free or wait, we denote by $C^{Yield}(t)$ the set of such vehicles: 
\begin{align}
C^{Yield}(t)&:=\{c^{j}\in{C\backslash{c^{i}}}:(x_{t}(c^{i})\leqslant{}x_{t}(c^{j}))\\&\wedge{}(\beta_{t}(c^{j})\lor{\alpha_{t}(c^{j})}=\beta_{t}(c^{i})))\wedge(s(t-h)=processing)\}.\nonumber
\end{align}
For HVs ahead of them, IVs conservatively consider an ``on'' turn light as an indication of a state being equal to $processing$.

\subsection{HV Lane Change and IV Following}
In this section we will address two issues, (i) how an HV changes lanes, and (ii) how an IV follows other vehicles. Throughout, we will assume that all IVs are in state $free$. In the next subsection we address how IVs change lanes. The constraints for lane change by HVs are loose. The safety-related set  i.e., the set of vehicles with respect to which an HV maintains a safe distance while following them only contains three vehicles: the nearest leading vehicle in the current lane, the nearest vehicle ahead of $c^{i}$ in the target lane, and the nearest  vehicle behind $c^{i}$ in its target lane.

\noindent\textbf{Rule for HVs in Multi-lane Traffic:} 

\noindent We suppose that an HV $c^{h}$ in state $wait$ and $free$ follows its nearest vehicle leading it in its current lane $c^{j2}:=\arg{\min}\,\{x_{t}(c): (x_{t}(c)>x_{t}(c^{h}))\wedge(\beta_{t}(c)=\beta_{t}(c^{h}))\}$. If an HV $c^{h}$ is in state $processing$ it follows both $c^{j2}$ and the front vehicle nearest to it in its target lane $c^{j1}:=\arg{\min}\,\{x_{t}(c): (x_{t}(c)\geq{}x_{t}(c^{h}))\wedge(\beta_{t}(c)=\alpha_{t}(c^{h}))\},$ with separation distance more than $d_{s}(v_{t-h}(c^{h}),a^{h}_{min})$ at time $t$ [3]. It can transit from state $wait$ to state $processing$ at time $t$ only if it finds that, for $c^{j1}$, $c^{j12}$, and the nearest vehicle behind it in its target lane $c^{i}=\arg{\max}\,\{x_{t}(c): (x_{t}(c)\leq{}x_{t}(c^{h}))\wedge(\beta_{t}(c)=\alpha_{t}(c^{h}))\}$, that the following conditions are simultaneously satisfied:

(i) $x_{t}(c^{j1})-x_{t}(c^{h})>{d_{s}(v_{t-h}(c^{h}),a^{h}_{min})}$

(ii) $x_{t}(c^{h})-x_{t}(c^{i})>{d_{s}(v_{t-h}(c^{i}),a^{h}_{min})}$.

This rule is reasonable because an HV such as $c^{h}$ does not need to differentiate between vehicle types or their states when identifying vehicles $c^{j1}$, $c^{j2}$ and $c^{i}$ before a lane change. Also, we note that HVs do not take into consideration any vehicles following them in their current lanes. Therefore, as will be designed in the protocol for following considered next, any IV which is following an HV in the same lane maintains enough space to prepare for possible lane change of that HV at any time. 

\noindent \textbf{Protocol for IVs in State Free or Wait When Following Other Vehicles:} 

\noindent Any IV $c^{i}$ in state $wait$ or $free$ follows its nearest lead vehicle $c^{j2}$ in its current lane with separation distances as follows:

(i) Suppose $C^{Yield}(t)=\emptyset$ If $c^{j2}\in{C_{I}}$, then $c^{i}$ follows $c^{j2}$ under MPC (5) with $D^{0}_{s}(v_{t-h},v^{Lead}_{t-h},a^{i}_{min})$ as separation distance at any time $t$; if $c^{j2}\in{C_{H}}$, then $c^{i}$ follows $c^{j2}$ under MPC (5) with $d_{s}(v_{t-h},a^{h}_{min})$ as separation distance at any time $t$;

(ii) If $C^{Yield}(t)\neq\emptyset$, then $c^{i}$ follows $c^{j2}=\arg{\min}\,\{x_{t}(c): ((x_{t}(c)>x_{t}(c^{h}))\wedge((\beta_{t}(c)=\beta_{t}(c^{h})))\lor(c\in{C^{Yield}}(t)))\}$ under MPC (5) with $d_{s}(v_{t-h},a^{h}_{min})$ as separation distance at any time $t$.

\noindent \textbf{Theorem 3.} Suppose that for some initial time $t$, the initial conditions in the protocols above are satisfied and all IVs are always in state $free$. Then the HVs are safe with respect to IVs if all the IVs move under the above protocol for following vehicles, and all HVs follow the designed rule.

\begin{proof}
First, consider the scenario when HVs are not changing a lane. Then each lane's safety is independent of the other lanes. Now, since $d_{s}(v_{t-h},a^{h}_{min})\geq D^{0}_{s}(v_{t-h},v^{Lead}_{t-h},a^{h}_{min})$ it follows that the system is safe by Theorem 2. Next, consider in addition a lane change of any HV $c^{h}\in{C^{H}}$. Since each HV follows and is followed at a  distance $d_{s}(v_{t-h},a^{h}_{min})$, the lane change is safe with respect to vehicles that do not change lanes, as proved in [3]. Therefore, no collision between IVs and HVs occur.
\end{proof}

\subsection{IV Lane Change}
We have shown in the previous  subsection that HVs stay safe with respect to IVs if all IVs are in state $free$. We also establish protocols for IVs in state $wait$ and $free$ to follow other vehicles. We now consider how lane change by IVs can be safely performed into the system in Section III.A. We first define a new MPC for governing the movement of an IV that is in state $processing$. 

\leftline{\textbf{MPC for Lane Change}}
\begin{align}
\underset{\mathbf{u}(0:N-1)}{\min}\,\hspace{0.1mm} &J(x_{t}, x_{t}^{f}, {\mathbf{u}(0:N-1)})\\
s.t.\hspace{4mm} 
&\mathbf{x}_{t+(k+1)h}=f(\mathbf{x}_{t+kh}, \mathbf{u}_{t+kh})\nonumber\\
&x_{t+(k+1)h}^{Lead}(c^{j})-x_{t+(k+1)h}\geqslant{D^{0}_{s}(v_{t+kh}, v_{t+kh}^{Lead}(c^{j}), a_{min})}\nonumber\\
&x_{t+(k+1)h}^{Lead}(c^{j})=x_{t+kh}^{Lead}(c^{j})+(v_{t+(k-1)h}^{Lead}(c^{j})+a_{min}h)h\nonumber\\
&c^{j}=\arg\underset{c\in C^{+}_{I_{1}}}{\min}\,\{x_{t}(c)-x_{t}+d_{s}(v_{t-h}(c),a_{min})\}\nonumber\\
&{x}_{t}(c^{l})-x_{t+(k+1)h}\geq{d_{s}(v_{t+kh},a_{min})}\nonumber\\
&c^{l}=\arg\underset{c\in C^{*}_{I_{2}}(t)\cup{C^{+}_{H}}}{\min}\,\{x_{t}(c)\}\nonumber\\
&a_{min}h\leqslant\Delta{v_{t+kh}}=v_{t+kh}-v_{t+(k-1)h}\leqslant{a_{max}h}\nonumber\\
&\theta_{min}\leqslant\theta_{t+(k+1)h}\leqslant{\theta_{max}}\nonumber\\
&v_{t+kh}\in\mathcal{V}^{k}_{t}=(\mathcal{V}_{t}(c^{j})\cap\mathcal{V}_{t}(c^{l}))^{k}_{t}\nonumber\\
&\omega_{t+kh}\in\Omega^{k}_{t}=(\Omega_{t}(c^{j})\cap\Omega_{t}(c^{l}))^{k}_{t},\nonumber
\end{align} 
where $k\in\{0,...,N-1\}$, $f(\mathbf{x}_{t+kh})$ as in (1), $\mathcal{V}_{t}(c^{j})$ is the constraint region for a vehicle in the single lane MPC (5) when $c^{i}$ is following $c^{j}$, $\mathcal{V}_{t}(c^{l})$ is the constraint region for the single lane MPC (10) in [3] when $c^{i}$ is following $c^{l}$. The velocity constraint region $\mathcal{V}_{t}$ for MPC (15) is an intersection of two velocity constraint regions. The angular velocity constraint reagion $\Omega_{t}$ for MPC (15) is constructed from $\mathcal{V}_{t}$ like [3]. We omit the identifier $c^{i}$ and assume all vehicles have the same parameter $a_{min}$ for simplicity. If the $y$ coordinate of the target lane is $W$, then the objective function can be: $J=\sum_{k=0}^{N-1}((y_{t+kh}-W)^{2}+\theta^{2}_{t+kh})$. 

$C^{*}_{I_{2}}(t)=C^{+}_{I_{2}}\cup (C^{+}_{I_{2}}(t)\cap \{c^{j}\in{C_{I}\backslash{c^{i}}}:(s(t-h)={processing})\})$ differs in different MPCs over different time slots during lane change, and contains vehicles ahead of $c^{i}$ which transit to state $processing$ after $c^{i}$ initiates its lane change. Constrained by condition (iii) of the lane change protocol, all such vehicles will not move before they ensure they can be followed safely by $c^{i}$. Since $C^{+}_{I_{1}}(t+kh)\backslash C^{+}_{I_{1}}(t)\subseteq C^{*}_{I_{2}}(t+kh)$ for some $k$, MPC for $c^{i}$ considers $C^{+}_{I_{1}}=C^{+}_{I_{1}}(t)$ unchanged during lane change. Similarly for $C^{+}_{I_{2}}=C^{+}_{I_{2}}(t)$. For $C_{H}^{+}(t)$, condition (iv) in the lane change protocol guarantees it is also unchanged during lane change.

\leftline{\textbf{Lane Change Protocol}} 
\noindent Suppose $c^{i}$ is in state $wait$, $\beta_{t}(c^{i})=0$, and $\alpha_{t}(c^{i})\in\{1,-1\}$. $c^{i}$ can transit to state $processing$ at time $t$ if the following conditions are satisfied:

(i) $x_{t}(c^{j})-x_{t}(c^{i})\geqslant{D^{0}_{s}(v_{t-h}(c^{i}),v_{t-h}(c^{j}),a_{min})}$

(ii) ${x}_{t}(c^{l})-x_{t}(c^{i})\geqslant{d_{s}(v_{t-h}(c^{i}),a_{min})}$

(iii) ${x}_{t}(c^{i})-{x}_{t}(c^{k})\geqslant{d_{s}(v_{t-h}(c^{k}),a_{min})}\forall\,\,c^{k}\in{C^{-}_{I}(t)}$

(iv)  ${x}_{t}(c^{i})-{x}_{t}(c^{k})\geqslant{v_{max}T+d_{s}(v_{max},a_{min})}\forall\,\,c^{k}\in{C^{-}_{H}(t)}$ for some $T>T_{min}(\mathcal{V}_{t})$\\
where $T_{min}(\mathcal{V}_{t})$ is the minimum time spent in lane change estimated based on velocity constraint region $\mathcal{V}_{t}$ if $\mathcal{V}_{t}\neq \emptyset$. The approach to give a feasible value for $T$ is illustrated in Lemma 2. 

Conditions (i)-(ii) when satisfied guarantee the non-emptiness of $\mathcal{V}_{t}$ for $t$-round MPC (15) by Theorem 1 and the proof of Theorem 2 in [3]. Condition (iii) when satisfied ensures that all vehicles in $C^{*}_{I_{2}}(t+kh)\backslash C^{*}_{I_{2}}(t)$ satisfy constraint (ii) at time $t+kh$ for any positive integer $k$. Hence, the existence of feasible solutions for the $t+kh$-round MPC is guaranteed. We note that condition (iv) is not independent with conditions (i)-(ii). Conditions (i)-(ii) being satisfied is a necessary condition for condition (iv) to hold. Condition (iv) when satisfied ensures that lane change of IV occurs either behind HVs with safety ensured by those IVs or in front of HVs. When it happens in front of HVs, those IVs ensure they can finish the lane change before HVs are within range $d_{s}(v_{t-h},a_{min})$.

\noindent \textbf{Lemma 1.} Suppose all IVs in state $wait$ follow the lane change protocol to transit from state $wait$ to state $processing$, and move under MPC (15) during lane change. Then all such IVs stay safe with respect to vehicles ahead of them during lane change.

\noindent \textbf{Lemma 2.} Suppose that at time $t$, conditions (i)-(iii) in the lane change protocol hold for $c^{i}$, $\theta_{t}=0$, and $c^{i}$ is in the middle line of lane $\beta_{t}(c^{i})$. Then the velocity constraint region is not empty and always contains deceleration with $a_{min}$ as a feasible solution. Thus if the achievable $y$-direction displacement of $c^{i}$ under this feasible velocity profile exceeds the lane width, we can bound $T_{min}(\mathcal{V}_{t})$ from above by $v_{t-h}/a_{min}$. Given the lane width $W_{l}$, if $d_{s}(v_{t-h}(c^{i}),a_{min})>3W_{l}/\sqrt{2}$, and
\begin{align}
d^{2}_{s}(v_{t-h}(c^{i}),a_{min})[1-\cos{\dfrac{d_{s}(v_{t-h}(c^{i}),a_{min})}{h}}]>3W^{2}_{l},
\end{align}
where the turning radius by $\rho=d^{2}_{s}(v_{t-h}(c^{i}),a_{min})/3W_{l}>3W_{l}/2$, then $v_{t-h}(c^{i})/a_{min}$ is an achievable upper bound for $T_{min}(\mathcal{V}_{t}),$ and can be used as the value of $T$ in condition (iv) of the lane change protocol.

\noindent \textbf{Theorem 4.} Suppose that at an initial time $t$, no IVs or HVs are in state $processing$, and the conditions in the protocol for following, and the rule for HVs in multi-lane traffic hold. Suppose also that: (i) IVs in state $free$ or state $wait$ move under the protocol for following, (ii) IVs follow the lane change protocol to transit from state $wait$ to $proceessing$, (iii) IVs in state $processing$ follow MPC (15), (iv) HVs follow the rule for HVs in multi-lane traffic. Then the HVs in this mixed
transportation system are safe with respect to IVs at any time
after $t$.

Platoon join and split at tail or head in multi-lane are handled similarly to single lane: retain all constraints to ensure safety and adjust objective function as listed in Table I. A new issue is split or join in the middle, which is a combination of lane change and split or join at tail or head. Thus, we treat a formed platoon no differently from an IV when dealing with inter-platoon safety and all above conclusions follow.
\section{SIMULATION RESULTS}
\begin{figure*}[thpb]
      \centering
      \includegraphics[scale=0.59]{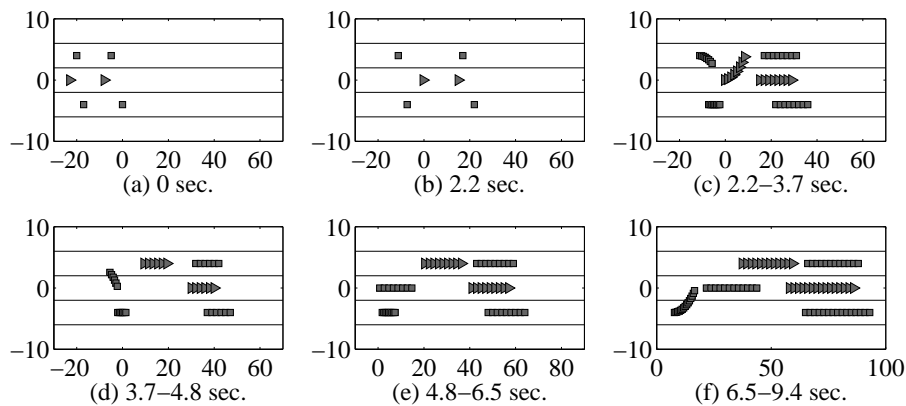}
      \vspace{-3mm}
      \caption{Simulation of a Multi-lane Traffic with Lane Change}
\end{figure*}
We present simulation results of the proposed system. We first simulate a mixed single lane with a gradually increasing percentage of IVs which are distributed uniformly along the single lane. Initially, all vehicles have the same speed $5m/s$, and have the same separation distance $5m$, which satisfy the initial conditions in Theorem 2, IVs follow their lead vehicles under Theorem 2 and HVs follow the rule for HVs in single lane traffic. We let consecutive IVs execute platoon join maneuvers and implement corresponding objective functions on their MPCs.
\begin{figure}[thpb]
      \centering
      \includegraphics[scale=0.4]{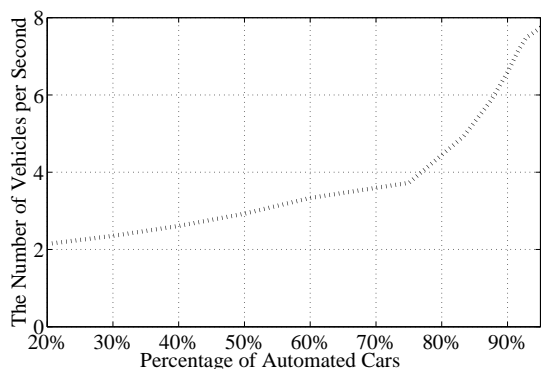}
      \vspace{-3mm}
      \caption{Throughput with Different Percentage of IVs}
\end{figure}

We count single lane throughput after several consecutive IVs have formed a platoon with intra-platoon spacing $2.5m$. The throughput of this single lane traffic with different percentages of IVs is shown in Fig. 4. We note that when the fraction of IVs is relatively large, the mixed system outperforms a system with only HVs. The simulation parameters are: $a^{i}_{min}=-8m/s^{2}$, $a^{h}_{min}=-6m/s^{2}$, $a_{max}=4m/s^{2}$, $h=0.01s$, $d_{min}=2m$ and $v_{max}= 42m/s$.

We next simulate multi-lane traffic with lane change. As an example, we consider six vehicles on a three-lane road as shown in Fig. 5. Triangles represent HVs and squares represent IVs. Lanes are numbered with 1, 0, and -1 from top to bottom. The target lane for the first vehicle in lane 1 is lane 0, for the first vehicle in lane 0 is lane 1, and for the first vehicle in lane -1 is lane 0. We adjust their objective functions in the MPC based on their targets. If the conditions in the lane change protocol  are satisfied for one such vehicle, it can initiate its lane change. Otherwise, it stays in state $wait$. From Fig. 5(a)-(c), we note that the first vehicles in lane 1 and lane -1 wait until the first vehicle in lane 0 passes them and is far enough from them, before initiating their lane change at a time when condition (iv) of lane change protocol works. Similarly, the rule for HVs in multi-lane traffic holds for the first vehicle in lane 0 in Fig. 5(c)-(d). Also, we note that in Fig. 5(e)-(f) the first vehicle in lane -1 transit from state $wait$ to state $processing$ when the first vehicle in lane 0 passes it and is far enough, where condition (i) in lane change protocol plays a role. The initial speed of the six vehicles is $10m/s$, and initial separation distance is $17m$.

\section{CONCLUDING REMARKS}
Anticipating a move towards increasing the fraction of automated vehicles in traffic systems, we have addressed the problem of provable safety in  mixed transportation systems. Motivated further by the goal of increasing throughput, we have also considered the adoption of strategies such as platooning by automated vehicles. For such systems we have established safety in both single lane scenarios as well as multi-lane scenarios with lane changes. The approach we follow considers looser models of human behavior and tighter models for automated vehicles. We employ a model predictive approach that allows incorporation of constraints of varying degree. Though MPC is a finite time methodology and safety is an infinite time property, we establish the safety of the resulting system. This approach can be extended to intelligent intersections, and will be described in a subsequent paper. 

\addtolength{\textheight}{-12cm}   





\end{document}